\documentclass[preprint,pra,showpacs]{revtex4}%
\usepackage{amsmath}
\usepackage{graphicx}
\usepackage{amsfonts}
\usepackage{amssymb}%
\newtheorem{theorem}{Theorem}
\newtheorem{acknowledgement}[theorem]{Acknowledgement}

\begin{document}
\preprint{ }
\title{Canonical and kinetic forms of the electromagnetic momentum in an \emph{ad
hoc} quantization scheme for a dispersive dielectric}
\author{J. C. Garrison}
\author{R. Y. Chiao}
\affiliation{{\small Department of Physics, University of California, Berkeley, California
94720}}

\begin{abstract}
An \emph{ad hoc} quantization scheme for the electromagnetic field in a weakly
dispersive, transparent dielectric leads to the definition of canonical and
kinetic forms for the momentum of the electromagnetic field in a dispersive
medium. The canonical momentum is uniquely defined as the operator that
generates spatial translations in a uniform medium, but the quantization
scheme suggests two possible choices for the kinetic momentum operator,
corresponding to the Abraham or the Minkowski momentum in classical
electrodynamics. Another implication of this procedure is that a wave packet
containing a single dressed photon travels at the group velocity through the
medium. The physical significance of the canonical momentum has already been
established by considerations of energy and momentum conservation in the
atomic recoil due to spontaneous emission, the Cerenkov effect, the Doppler
effect, and phase matching in nonlinear optical processes. In addition, the
data of the Jones and Leslie radiation pressure experiment is consistent with
the assignment of one $\hbar\mathbf{k}$ unit of canonical momentum to each
dressed photon. By contrast, experiments in which the dielectric is rigidly
accelerated by unbalanced electromagnetic forces require the use of the
Abraham momentum.

\end{abstract}

\maketitle

\section{\label{&introduction}Introduction}

In classical electrodynamics, a medium is traditionally described by its
macroscopic linear susceptibility. The long history and great utility of this
phenomenological method have inspired a substantial body of work aimed at
devising a similar description for the quantized electromagnetic field in a
dielectric medium
\cite{Drummond1990,Glauber-Lewenstein1991,Huttner1992,Matloob1995,Gruner1996,Loudon2002}%
. This has proven to be a difficult and subtle task.

A useful \emph{ad hoc} scheme for the quantization of the electromagnetic
field in a dispersive dielectric has been independently suggested by
Loudon\cite{Blow1990} and Milonni \cite{Milonni1995}. In the present paper we
will use Milonni's version of this scheme. This simple and plausible
formulation leads in a natural way to the definition of several forms of
electromagnetic momentum; a \textquotedblleft canonical\textquotedblright%
\ momentum\ associated with spatial translations, and two \textquotedblleft
kinetic\textquotedblright\ momenta\ that result from quantizing the familiar
Abraham or Minkowski momenta of classical electrodynamics. We shall see that
all of these operators can be physically meaningful, but that they have
different domains of applicability.

The existence of more than one form of momentum may seem surprising, but there
is an analogous situation in semiclassical electrodynamics. In the
nonrelativistic limit, the kinetic energy part of the Hamiltonian for this
problem is%
\begin{equation}
H=\frac{1}{2m}\left(  \mathbf{p}-e\boldsymbol{\mathcal{A}} \right)  ^{2},
\label{nrham}%
\end{equation}
where $m$ is the mass, $e$ is the charge, $\boldsymbol{\mathcal{A}} $ is the
classical vector potential (we shall use calligraphic symbols for all
classical variables), and
\begin{equation}
\mathbf{p}=\frac{\hbar}{i}\mathbf{\nabla} \label{canonical-mom}%
\end{equation}
is the \textquotedblleft canonical\textquotedblright\ momentum
\cite{Feynman1965}. The Heisenberg equation of motion $i\hslash d\mathbf{r}%
/dt=\left[  \mathbf{r},H\right]  $ shows that the velocity operator
$\mathbf{v}=d\mathbf{r}/dt$ is given by%
\begin{equation}
m\mathbf{v}=\mathbf{p}-e\boldsymbol{\mathcal{A}} , \label{kinetic-mom}%
\end{equation}
and this defines the \textquotedblleft kinetic\textquotedblright\ momentum
$m\mathbf{v}$.

The kinetic momentum in (\ref{kinetic-mom}) evidently has the expected
classical limit, i.e., the product of mass and velocity, but it does not serve
as the generator of spatial translations. To see this, we note that spatial
translations along different axes commute, so that the corresponding
generators must also commute. An explicit calculation using (\ref{kinetic-mom}%
) yields%
\begin{equation}
\left[  mv_{i},mv_{j}\right]  =i\hslash e\epsilon_{ijk}\mathcal{B}_{k}\neq0,
\end{equation}
where $\boldsymbol{\mathcal{B}} =\mathbf{\nabla\times}\boldsymbol{\mathcal{A}}
$ is the magnetic field, and the Einstein summation convention is used for
repeated vector indices. This shows that $m\mathbf{v}$ cannot be the generator
of spatial translations for $\boldsymbol{\mathcal{B}} \neq0$. On the other
hand, it is well known that the canonical momentum $\mathbf{p}$ in
(\ref{canonical-mom}) is the operator that generates spatial translations, but
solving (\ref{kinetic-mom}) for $\mathbf{p}$ shows that it does not have the
expected classical limit. Thus both the canonical and kinetic momenta are
physically meaningful, but\ they play distinct roles in the theory.

In the following sections, we shall see that Milonni's quantization scheme
leads to an analogous situation. In the electromagnetic case there is a unique
``canonical''\ momentum operator $\mathbf{P}_{can}$ that generates spatial
translations, but there are at least two possibilities for the kinetic
momentum. This peculiar situation is related to the long standing controversy
in classical electrodynamics regarding the ``correct''\ definition of the
electromagnetic momentum density in a medium \cite{Landau1984},
\cite{Ginzburg1989}. The traditional contenders for this title are the
Abraham,%
\begin{equation}
\mathbf{g}_{A}\left(  \mathbf{r},t\right)  =\frac{\left\langle
\boldsymbol{\mathcal{E}} \left(  \mathbf{r},t\right)  \mathbf{\times
}\boldsymbol{\mathcal{H}} \left(  \mathbf{r},t\right)  \right\rangle }{c^{2}},
\label{Abraham}%
\end{equation}
and the Minkowski,%
\begin{equation}
\mathbf{g}_{M}\left(  \mathbf{r},t\right)  =\left\langle
\boldsymbol{\mathcal{D}} \left(  \mathbf{r},t\right)  \mathbf{\times
}\boldsymbol{\mathcal{B}} \left(  \mathbf{r},t\right)  \right\rangle ,
\label{Minkowski}%
\end{equation}
forms of the momentum density, where $\left\langle \cdots\right\rangle $
indicates an average over the period of the carrier wave. At present there
seems to be a fairly strong consensus that the Abraham form is to be preferred
for the electromagnetic momentum density
\cite{Ginzburg1989,Jackson1999a,Jackson1999b,Lai1979,Lai1981,Gordon1973}, but
new proposals continue to appear. In the work of Obukhov and Hehl
\cite{Obukhov2003}, for example, the energy momentum tensor is automatically
symmetric, and it leads to the momentum density $\mathbf{g}_{OH}\left(
\mathbf{r},t\right)  =\epsilon_{0}\left\langle \boldsymbol{\mathcal{E}}
\left(  \mathbf{r},t\right)  \mathbf{\times}\boldsymbol{\mathcal{B}} \left(
\mathbf{r},t\right)  \right\rangle $. In the present paper we only consider
nonmagnetic materials $\left(  \mu=\mu_{0}\right)  $, for which $\mathbf{g}%
_{OH}\left(  \mathbf{r},t\right)  =\mathbf{g}_{A}\left(  \mathbf{r},t\right)
$, but it would be interesting to see an application of the Obukhov-Hehl
approach to dispersive dielectrics. As pointed out by Loudon \cite{Loudon2002}%
, the various forms of the momentum are potentially useful in different
contexts. It should also be noted that Brevik \cite{Brevik1979} has argued
that there is no unique solution to the problem of identifying the
``true''\ electromagnetic energy-momentum tensor, since there is no unique
prescription for the separation of the total energy-momentum tensor into a
field part and a matter part. DeGroot and Suttorp \cite{DeGroot-Suttorp1972}
have pointed out that the problem of deriving the forms of the energy, the
linear momentum, and the angular momentum for polarized media cannot be solved
as long as macroscopic arguments are utilized; microscopic arguments starting
from statistical mechanics are necessary.

In Section \ref{&quantization}, we present Milonni's procedure for the
quantization of electromagnetic fields in a weakly dispersive, transparent
dielectric medium. In Section \ref{&canonical}, we show that identifying the
total electromagnetic momentum with the uniquely defined generator of spatial
translations is equivalent to assuming that a photon with wave vector
$\mathbf{k}$ has momentum $\hslash\mathbf{k}$, just as in the vacuum. The
importance of the generator of spatial translations in this connection was
previously noted by Brevik and Lautrop \cite{Brevik1970}, but their work was
limited to nondispersive materials. In Section \ref{&kinetic}, we show that
quantization of the familiar Abraham and Minkowski versions of the total
electromagnetic momentum leads to alternative suggestions for the form of the
single-photon momentum. In Section \ref{&experiment} we discuss experimental
tests of the predictions of this quantization method.

\section{\label{&quantization}Quantization in a dispersive dielectric}

Milonni's method of quantization of the electromagnetic field in a weakly
dispersive, transparent dielectric has the twin virtues of simplicity and
agreement with the much more elaborate formalisms developed in some of the
other references cited in the introduction. This approach is directly based on
the approximations used in the classical theory, so we begin by considering a
classical field described by the vector potential,%
\begin{equation}
\boldsymbol{\mathcal{A}} \left(  \mathbf{r},t\right)  =%
\boldsymbol{\mathcal{A}}%
^{\left(  +\right)  }\left(  \mathbf{r},t\right)  +CC, \label{field1}%
\end{equation}
where the analytic signal $%
\boldsymbol{\mathcal{A}}%
^{\left(  +\right)  }\left(  \mathbf{r},t\right)  $ is given by
\begin{equation}%
\boldsymbol{\mathcal{A}}%
^{\left(  +\right)  }\left(  \mathbf{r},t\right)  =\int\frac{d^{3}k}{\left(
2\pi\right)  ^{3}}\sum_{s}\mathcal{A}_{s}\left(  \mathbf{k}\right)
\mathbf{e}_{s}\left(  \mathbf{k}\right)  e^{i\left(  \mathbf{k\cdot r}%
-\omega\left(  k\right)  t\right)  }. \label{propclasclsol}%
\end{equation}
For the quasimonochromatic fields of interest, the power spectrum, $\left\vert
\mathcal{A}_{s}\left(  \mathbf{k}\right)  \right\vert ^{2}$ is concentrated at
a particular frequency $\omega_{0}$ with spectral width $\Delta\omega
<<\omega_{0}$. The medium is assumed to be weakly dispersive with respect to
this wave packet, i.e.,%
\begin{equation}
\Delta n=\Delta\omega\left\vert \left(  \frac{\partial n\left(  \omega\right)
}{\partial\omega}\right)  _{\omega=\omega_{0}}\right\vert <<\left\vert
n\left(  \omega_{0}\right)  \right\vert . \label{wkdis02}%
\end{equation}

For classical fields satisfying (\ref{field1})-(\ref{wkdis02}) the effective
energy is \cite{Jackson1999a}
\begin{equation}
\mathcal{U}_{em}=\frac{d\left[  \omega_{0}\epsilon\left(  \omega_{0}\right)
\right]  }{d\omega_{0}}\frac{1}{2}\int d^{3}r\left\langle
\boldsymbol{\mathcal{E}}%
^{2}\left(  \mathbf{r},t\right)  \right\rangle +\frac{1}{2\mu_{0}}\ \int
d^{3}r\left\langle
\boldsymbol{\mathcal{B}}%
^{2}\left(  \mathbf{r},t\right)  \right\rangle ,
\end{equation}
where $\left\langle \cdots\right\rangle $ denotes an average over the carrier
period $2\pi/\omega_{0}$. By using (\ref{propclasclsol}) one can carry out the
volume integrals to get%
\begin{equation}
\mathcal{U}_{em}=\int\frac{d^{3}k}{\left(  2\pi\right)  ^{3}}\sum_{s}\left\{
\omega^{2}\left(  k\right)  \frac{d\left[  \omega_{0}\epsilon\left(
\omega_{0}\right)  \right]  }{d\omega_{0}}+\frac{k^{2}}{\mu_{0}}\right\}
\left\vert \mathcal{A}_{s}\left(  \mathbf{k}\right)  \right\vert ^{2},
\label{safestep}%
\end{equation}
and the narrow width of the power spectrum allows this to be rewritten this in
the more suggestive form%
\begin{equation}
\mathcal{U}_{em}=\int\frac{d^{3}k}{\left(  2\pi\right)  ^{3}}\sum_{s}\left\{
\omega^{2}\left(  k\right)  \frac{d\left[  \omega\left(  k\right)
\epsilon\left(  \omega\left(  k\right)  \right)  \right]  }{d\omega\left(
k\right)  }+\frac{k^{2}}{\mu_{0}}\right\}  \left\vert \mathcal{A}_{s}\left(
\mathbf{k}\right)  \right\vert ^{2}. \label{dangerstep}%
\end{equation}
This step is both dangerous and useful. The danger comes from the apparent
generality of (\ref{dangerstep}), which might lead one to forget that it was
derived for a quasimonochromatic field. The utility comes from the observation
that this expression is also valid for a superposition of quasimonochromatic
fields, provided that the differences between the carrier frequencies are
large compared to the spectral widths of the individual wave packets. In this
situation we shall say that the total field is \textquotedblleft
quasimultichromatic\textquotedblright. With these caveats held firmly in mind,
we use the relation $\epsilon\left(  \omega\left(  k\right)  \right)
=\epsilon_{0}n^{2}\left(  \omega\left(  k\right)  \right)  $ to rewrite
(\ref{dangerstep}) as%
\begin{equation}
\mathcal{U}_{em}=2\epsilon_{0}\int\frac{d^{3}k}{\left(  2\pi\right)  ^{3}}%
\sum_{s}\frac{\omega^{2}\left(  k\right)  n\left(  k\right)  }{\left(
v_{gr}\left(  k\right)  /c\right)  }\left\vert \mathcal{A}_{s}\left(
\mathbf{k}\right)  \right\vert ^{2}, \label{cleng}%
\end{equation}
where%
\begin{equation}
v_{gr}\left(  k\right)  =\frac{d\omega}{dk}=\frac{c}{n\left(  k\right)
+\omega\left(  k\right)  \left(  dn/d\omega\right)  _{k}}, \label{vgrp}%
\end{equation}
is the group velocity and $v_{ph}\left(  k\right)  =c/n\left(  k\right)  $ is
the phase velocity

The next step is to express the energy as the sum of energies $\hslash
\omega\left(  k\right)  $ of radiation oscillators. To this end we define new
amplitudes $\alpha_{s}\left(  \mathbf{k}\right)  $ by the rule
\begin{equation}
\mathcal{A}_{s}\left(  \mathbf{k}\right)  =\sqrt{\frac{\hslash\left(
v_{gr}\left(  k\right)  /c\right)  }{2\epsilon_{0}n\left(  k\right)
\omega\left(  k\right)  }}\alpha_{s}\left(  \mathbf{k}\right)  , \label{anorm}%
\end{equation}
so that $\left\vert \alpha_{s}\left(  \mathbf{k}\right)  \right\vert ^{2}$
(with dimensions $L^{3}$) is a $\mathbf{k}$-space density. The resulting
expression,
\begin{equation}
\mathcal{U}_{em}=\int\frac{d^{3}k}{\left(  2\pi\right)  ^{3}}\sum_{s}%
\hslash\omega\left(  k\right)  \left\vert \alpha_{s}\left(  \mathbf{k}\right)
\right\vert ^{2}, \label{engdis02}%
\end{equation}
for the total energy opens the way to the standard quantization rule%
\begin{equation}
\alpha_{s}\left(  \mathbf{k}\right)  \rightarrow a_{s}\left(  \mathbf{k}%
\right)  ,\ \ \alpha_{s}^{\ast}\left(  \mathbf{k}\right)  \rightarrow
a_{s}^{\dagger}\left(  \mathbf{k}\right)  , \label{qrule}%
\end{equation}
where the operators $a_{s}\left(  \mathbf{k}\right)  $ and $a_{s}^{\dagger
}\left(  \mathbf{k}\right)  $ satisfy the canonical commutation relations,
\begin{equation}
\left[  a_{s}\left(  \mathbf{k}\right)  ,a_{s^{\prime}}^{\dagger}\left(
\mathbf{k}^{\prime}\right)  \right]  =\left(  2\pi\right)  ^{3}\delta
_{ss^{\prime}}\delta^{(3)}\left(  \mathbf{k}-\mathbf{k}^{\prime}\right)  .
\end{equation}
In this scheme the Hamiltonian and the positive-frequency part of the field
are respectively given by%
\begin{equation}
H_{em}=\int\frac{d^{3}k}{\left(  2\pi\right)  ^{3}}\sum_{s}\hslash
\omega\left(  k\right)  a_{s}^{\dagger}\left(  \mathbf{k}\right)  a_{s}\left(
\mathbf{k}\right)  , \label{hamdis2}%
\end{equation}
and%
\begin{equation}
\mathbf{A}^{\left(  +\right)  }\left(  \mathbf{r}\right)  =\int\frac{d^{3}%
k}{\left(  2\pi\right)  ^{3}}\sum_{s}\sqrt{\frac{\hslash v_{gr}\left(
k\right)  }{2\epsilon_{0}n\left(  k\right)  \omega\left(  k\right)  c}}%
a_{s}\left(  \mathbf{k}\right)  \mathbf{e}_{s}\left(  \mathbf{k}\right)
e^{i\mathbf{k\cdot r}}\text{.} \label{aandis2}%
\end{equation}
The excitations created by $a_{s}^{\dagger}\left(  \mathbf{k}\right)  $ are
quasiparticles that contain some admixture of electromagnetic and atomic
degrees of freedom, i.e., they are \textquotedblleft dressed\textquotedblright%
\ photons. This is in the spirit of Einstein's original model of light quanta
in the vacuum, since each dressed photon carries energy $\hslash\omega\left(
k\right)  $ according to (\ref{hamdis2}). Furthermore, one can show that the
appearance of the group velocity in the normalization factor in (\ref{aandis2}%
) guarantees that a single-photon wave packet, propagating at the group
velocity, carries the energy $\hslash\omega\left(  k\right)  $ associated with
the carrier wave.

The classical quasimultichromatic approximation implies that a plot of the
power spectrum $\left\vert \alpha_{s}\left(  \mathbf{k}\right)  \right\vert
^{2}$ must consist of a set of narrow peaks centered on the carrier
frequencies of the wave packets making up the classical field, but this
condition makes no sense when applied to the operator $a_{s}^{\dagger}\left(
\mathbf{k}\right)  a_{s}\left(  \mathbf{k}\right)  $. In the quantum theory
this kind of information is carried by the states, so we need to choose a
subspace $\mathfrak{H}_{qm}$ of the total electromagnetic Fock space that
corresponds to the classical quasimultichromatic field \cite{Deutsch1991b}.
The number states
\begin{equation}
\left\vert \underline{n}\right\rangle \equiv\left\vert n_{s_{1}}\left(
\mathbf{k}_{1}\right)  ,n_{s_{2}}\left(  \mathbf{k}_{2}\right)
,...\right\rangle ,
\end{equation}
that satisfy%
\begin{equation}
a_{s_{j}}^{\dagger}\left(  \mathbf{k}_{j}\right)  a_{s_{j}}\left(
\mathbf{k}_{j}\right)  \left\vert \underline{n}\right\rangle =n_{s_{j}}\left(
\mathbf{k}_{j}\right)  \left\vert \underline{n}\right\rangle ,
\end{equation}
provide a basis for the entire Fock space, so the subspace $\mathfrak{H}_{qm}$
can be defined as the set of all linear combinations of number states
satisfying the condition that $n_{s}\left(  \mathbf{k}\right)  =0$ unless
$\omega\left(  k\right)  $ lies in a narrow band centered on one of the
carrier frequencies. The operator expressions (\ref{hamdis2}) and
(\ref{aandis2}) are valid only when applied to state vectors in $\mathfrak{H}%
_{qm}$.

\section{\label{&canonical}Canonical momentum}

The quantization scheme presented in Section II involves the following
assumptions: (a) The medium can only respond through the electronic
polarization of the atoms; no center-of-mass motion is allowed. (b) The
material response is spatially homogeneous, at least on the scale of optical
wavelengths. (c) The medium is isotropic. Assumption (c) (which is valid for
vapors, liquids, and glasses) justifies the use of a scalar dielectric
function. The quantization scheme can be generalized to crystals by using a
dielectric tensor instead.

The combination of assumptions (a) and (b) implies that the positional and
inertial degrees of freedom of the constituent atoms are irrelevant in this
model. As a consequence of these assumptions, the generator, $\mathbf{P}%
_{can}$, of spatial translations is completely defined by its action on the
field operators,%
\begin{equation}
\left[  A_{j}^{\left(  +\right)  }\left(  \mathbf{r}\right)  ,\mathbf{P}%
_{can}\right]  =\frac{\hslash}{i}\mathbf{\nabla}A_{j}^{\left(  +\right)
}\left(  \mathbf{r}\right)  .
\end{equation}
Using the expansion (\ref{aandis2}) to evaluate both sides leads to
\begin{equation}
\left[  a_{s}\left(  \mathbf{k}\right)  ,\mathbf{P}_{can}\right]
=\hslash\mathbf{k}a_{s}\left(  \mathbf{k}\right)  .
\end{equation}
The operator
\begin{equation}
\mathbf{P}_{can}=\int\frac{d^{3}k}{\left(  2\pi\right)  ^{3}}\sum_{s}%
\hslash\mathbf{k}a_{s}^{\dagger}\left(  \mathbf{k}\right)  a_{s}\left(
\mathbf{k}\right)  , \label{momdis}%
\end{equation}
obviously satisfies this condition. Any alternative form $\mathbf{P}%
_{can}^{\prime}$ would have to satisfy $\left[  a_{s}\left(  \mathbf{k}%
\right)  ,\mathbf{P}_{can}^{\prime}-\mathbf{P}_{can}\right]  =0$ for all modes
$\mathbf{k}s$, which is only possible if the operator $\mathbf{Z}%
\equiv\mathbf{P}_{can}^{\prime}-\mathbf{P}_{can}$ is actually a $c$-number. In
this case $\mathbf{Z}$ can be set to zero, for example by assuming that the
vacuum state is an eigenstate of $\mathbf{P}_{can}$ with zero eigenvalue, or
equivalently that the vacuum state is invariant under spatial translations. By
analogy with (\ref{canonical-mom}) we will call $\mathbf{P}_{can}$ the
``canonical momentum''\ of the field. From (\ref{momdis}) we then see that a
photon with wave vector $\mathbf{k}$ propagating in a dispersive medium is
assigned the momentum $\hslash\mathbf{k}$, just as in the vacuum.

One physical justification for the interpretation of $\mathbf{P}_{can}$ as a
form of electromagnetic momentum is provided by the empirical fact that this
$\hslash\mathbf{k}$-type of momentum is conserved in nonlinear optical
processes, such as spontaneous parametric down-conversion. In this process an
initial photon with energy and momentum $\left(  \hbar\omega_{0}%
,\hbar\mathbf{k}_{0}\right)  $ spontaneously decays into two down-converted
photons with energies and momenta $\left(  \hbar\omega_{1},\hbar\mathbf{k}%
_{1}\right)  $ and $\left(  \hbar\omega_{2},\hbar\mathbf{k}_{2}\right)  $,
respectively, so as to conserve energy and canonical momentum through the
well-verified phase-matching conditions \cite{Saleh1991}
\begin{equation}
\hbar\omega_{0}=\hbar\omega_{1}+\hbar\omega_{2}\text{ and }\hbar\mathbf{k}%
_{0}=\hbar\mathbf{k}_{1}+\hbar\mathbf{k}_{2}\text{ .} \label{phasematch}%
\end{equation}
Further pieces of evidence are that the spontaneous emission of a photon with
wave vector $\mathbf{k}$ in the medium results in an atomic recoil momentum
$\mathbf{p}_{rec}=\hslash\mathbf{k}$, and that the Cerenkov and Doppler
effects are also simply explained by the assignment of a momentum
$\hbar\mathbf{k}$ to each emitted photon \cite{Ginzburg1989}%
,\cite{Ginzburg1973}.

An isotropic medium is invariant under continuous rotations, so an extension
of the above argument shows that the rotation generator\ $\mathbf{J}_{can}$ is
again entirely defined by its action on the fields
\begin{equation}
\left[  A_{j}^{\left(  +\right)  }\left(  \mathbf{r}\right)  ,\left(
\mathbf{J}_{can}\right)  _{i}\right]  =\left(  \mathbf{r\times}\frac{\hslash
}{i}\mathbf{\nabla}\right)  _{i}A_{j}^{\left(  +\right)  }\left(
\mathbf{r}\right)  +i\hslash\epsilon_{jik}A_{k}^{\left(  +\right)  }\left(
\mathbf{r}\right)  .
\end{equation}
Substituting (\ref{aandis2}) into this condition yields the commutator%
\begin{equation}
\left[  a_{s}\left(  \mathbf{k}\right)  ,\left(  \mathbf{J}_{can}\right)
_{i}\right]  =\frac{\hslash}{i}e_{sj}^{\ast}\left(  \mathbf{k}\right)  \left(
\mathbf{k\times}\frac{\partial}{\partial\mathbf{k}}\right)  _{i}\left\{
\sum_{r}a_{r}\left(  \mathbf{k}\right)  e_{rj}\left(  \mathbf{k}\right)
\right\}  +\hslash s\frac{k_{i}}{k}a_{s}\left(  \mathbf{k}\right)  ,
\end{equation}
and inspection shows that $\mathbf{J}_{can}$ is given by the standard form
\cite{Mandel1995}
\begin{equation}
\mathbf{J}_{can}=\int\frac{d^{3}k}{\left(  2\pi\right)  ^{3}}\left\{
\mathbf{a}_{j}^{\dagger}\left(  \mathbf{k}\right)  \left(  \hslash
\mathbf{k\times}\frac{1}{i}\frac{\partial}{\partial\mathbf{k}}\right)
_{i}\mathbf{a}_{j}\left(  \mathbf{k}\right)  +\frac{\mathbf{k}}{k}\sum
_{s}\hslash sa_{s}^{\dagger}\left(  \mathbf{k}\right)  a_{s}\left(
\mathbf{k}\right)  \right\}  , \label{angmomdis}%
\end{equation}
where%
\begin{equation}
\mathbf{a}\left(  \mathbf{k}\right)  =\sum_{s}a_{s}\left(  \mathbf{k}\right)
\mathbf{e}_{s}\left(  \mathbf{k}\right)  .
\end{equation}

\section{\label{&kinetic}Kinetic momenta}

The quantization scheme we are using starts with the standard classical
expression for the electromagnetic energy in a dispersive dielectric, so it
would seem natural to construct the operators for momentum and angular
momentum by applying the same quantization rule (\ref{qrule}) to the
appropriate classical expressions. Since it is precisely the identification of
the appropriate expressions that is disputed in the Abraham \emph{vs
}Minkowski controversy, we must consider both possibilities. Integrating
(\ref{Abraham}) and (\ref{Minkowski}) over all space leads to the rival
expressions%
\begin{equation}
\boldsymbol{\mathcal{P}} _{A}=\int d^{3}r\mathbf{g}_{A}\left(  \mathbf{r}%
,t\right)  =\int\frac{d^{3}k}{\left(  2\pi\right)  ^{3}}\frac{S\left(
\mathbf{k}\right)  }{c^{2}}\frac{\mathbf{k}}{k}, \label{abrahamtot}%
\end{equation}
and
\begin{equation}
\boldsymbol{\mathcal{P}} _{M}=\int d^{3}r\mathbf{g}_{M}\left(  \mathbf{r}%
,t\right)  =\int\frac{d^{3}k}{\left(  2\pi\right)  ^{3}}\frac{S\left(
\mathbf{k}\right)  }{v_{ph}^{2}\left(  k\right)  }\frac{\mathbf{k}}{k},
\label{minkowskitot}%
\end{equation}
for the total momentum, where%
\begin{equation}
S\left(  \mathbf{k}\right)  =2\epsilon_{0}c^{2}k\omega\left(  k\right)
\sum_{s}\left|  \mathcal{A}_{s}\left(  \mathbf{k}\right)  \right|  ^{2}%
\end{equation}
is the time-averaged magnitude of the Poynting flux. Applying the quantization
rule (\ref{qrule}) to $\boldsymbol{\mathcal{P}} _{A}$ and
$\boldsymbol{\mathcal{P}} _{M}$ produces the operators%
\begin{align}
\mathbf{P}_{A}  &  =\int\frac{d^{3}k}{\left(  2\pi\right)  ^{3}}\sum
_{s}\hslash\omega\left(  k\right)  \frac{v_{gr}\left(  k\right)  }{c^{2}}%
\frac{\mathbf{k}}{k}a_{s}^{\dagger}\left(  \mathbf{k}\right)  a_{s}\left(
\mathbf{k}\right)  ,\nonumber\\
&  =\int\frac{d^{3}k}{\left(  2\pi\right)  ^{3}}\sum_{s}\frac{v_{gr}\left(
k\right)  v_{ph}\left(  k\right)  }{c^{2}}\hslash\mathbf{k\ }a_{s}^{\dagger
}\left(  \mathbf{k}\right)  a_{s}\left(  \mathbf{k}\right)  ,
\label{AbrahamTotMom2}%
\end{align}
and%
\begin{align}
\mathbf{P}_{M}  &  =\int\frac{d^{3}k}{\left(  2\pi\right)  ^{3}}\sum
_{s}\hslash\omega\left(  k\right)  n^{2}\left(  k\right)  \frac{v_{gr}\left(
k\right)  }{c^{2}}\frac{\mathbf{k}}{k}a_{s}^{\dagger}\left(  \mathbf{k}%
\right)  a_{s}\left(  \mathbf{k}\right)  ,\nonumber\\
&  =\int\frac{d^{3}k}{\left(  2\pi\right)  ^{3}}\sum_{s}\frac{v_{gr}\left(
k\right)  }{v_{ph}\left(  k\right)  }\hslash\mathbf{k\ }a_{s}^{\dagger}\left(
\mathbf{k}\right)  a_{s}\left(  \mathbf{k}\right)  . \label{MinkowskiTotMom2}%
\end{align}
Conversely, the classical limit of $\mathbf{P}_{A}\left(  \mathbf{P}%
_{M}\right)  $ is $\boldsymbol{\mathcal{P}} _{A}\left(
\boldsymbol{\mathcal{P}} _{M}\right)  $. Comparing these expressions to the
canonical momentum (\ref{momdis}) shows that---just as for the kinetic
momentum in (\ref{kinetic-mom})---neither of these kinetic momentum operators
is the generator of spatial translations.

Since $a_{s}^{\dagger}\left(  \mathbf{k}\right)  a_{s}\left(  \mathbf{k}%
\right)  $ is the number operator for photons in the $\mathbf{k}s$-mode, the
expressions (\ref{AbrahamTotMom2}) and (\ref{MinkowskiTotMom2}) imply that a
single dressed photon in a dispersive dielectric has the momentum
\begin{equation}
\mathbf{p}_{A}=\frac{v_{gr}\left(  k\right)  }{cn\left(  k\right)  }%
\hslash\mathbf{k}=\frac{v_{gr}\left(  k\right)  v_{ph}\left(  k\right)
}{c^{2}}\hslash\mathbf{k,} \label{AbrahamPhotonMom}%
\end{equation}
for the Abraham form, and
\begin{equation}
\mathbf{p}_{M}=\frac{n\left(  k\right)  v_{gr}\left(  k\right)  }{c}%
\hslash\mathbf{k}=\frac{v_{gr}\left(  k\right)  }{v_{ph}\left(  k\right)
}\hslash\mathbf{k} \label{MinkowskiPhotonMom}%
\end{equation}
for the Minkowski form. It has been experimentally verified
\cite{steinbergPRL1992} that a single-photon wave packet propagates at the
group velocity $v_{gr}\left(  k\right)  <c$ in a passive, transparent medium,
such as glass. Roughly speaking, the peak of the wave packet indicates the
most likely ``position''\ of the photon, when it is regarded as a particle.
This suggests that the dressed photon might be regarded as a relativistic
particle with velocity $v_{gr}\left(  k\right)  $, and this would in turn lead
to the definition of an effective mass as $\left|  \mathbf{p}\right|  /v_{gr}$.

In the Abraham picture, a dressed photon in a dispersive medium has the
effective mass
\begin{equation}
m_{\text{ }A}^{eff}=\frac{p_{A}}{v_{gr}\left(  k\right)  }=\frac{\hslash
\omega\left(  k\right)  }{c^{2}}. \label{m-eff-A}%
\end{equation}
This is what is sometimes called the relativistic mass, and should not be
confused with the rest mass. Thus the single-quantum energy $\hslash
\omega\left(  k\right)  $ determines the relativistic inertial mass of the
dressed photon. This is consistent with Planck's law of inertia for
electromagnetic energy \cite{Planck1908}, which states that for any closed
system containing a dielectric, the ratio of the momentum density to the
energy flux is given by $1/c^{2}$. Planck's law of inertia was formulated
classically for nondispersive media, but this definition of the effective mass
generalizes it to the quantum level, and includes dispersive dielectrics. Thus
we interpret Planck's law of inertia to mean that each dressed photon
contributes an inertial mass, given by (\ref{m-eff-A}), to a blackbody cavity
which is filled with a uniform dielectric, and which is undergoing rigid-body
acceleration. We should also note that Planck's law of inertia is
automatically satisfied by the Obukhov-Hehl form of the energy-momentum tensor.

In the Minkowski picture, the dressed photon propagating inside the dielectric
medium possesses an effective mass%
\begin{equation}
m_{M}^{eff}=\frac{p_{M}}{v_{gr}\left(  k\right)  }=\frac{\hslash\omega\left(
k\right)  }{v_{ph}^{2}\left(  k\right)  }=n^{2}\left(  k\right)  \frac
{\hslash\omega\left(  k\right)  }{c^{2}},
\end{equation}
which differs from the Abraham expression by the extra factor $n^{2}\left(
k\right)  $ in the numerator.

Comparing the three expressions (\ref{AbrahamTotMom2}),
(\ref{MinkowskiTotMom2}), and (\ref{momdis}) for $\mathbf{P}_{A}$,
$\mathbf{P}_{M}$, and $\mathbf{P}_{can}$, respectively, shows that
$\mathbf{P}_{A}=\mathbf{P}_{M}$ can only hold if $n^{2}\left(  k\right)  =1$,
i.e., for the vacuum, and that $\mathbf{P}_{A}=\mathbf{P}_{can}$ is only
possible in the unlikely special case that $v_{gr}\left(  k\right)
v_{ph}\left(  k\right)  =c^{2}$. On the other hand, equality between
$\mathbf{P}_{can}$ and $\mathbf{P}_{M}$ occurs for any nondispersive medium,
i.e., whenever there is a range of frequencies for which $\omega dn\left(
\omega\right)  /d\omega<<n\left(  \omega\right)  $. In this case the phase and
group velocities coincide, and $\mathbf{P}_{can}=\mathbf{P}_{M}$. This
situation occurs automatically in the low-frequency or static limit
$\omega\rightarrow0$, since (\ref{vgrp}) shows that $v_{gr}\left(  0\right)
=v_{ph}\left(  0\right)  $. This is a good approximation for dielectrics in
the low-frequency limit, as was pointed out by Gordon \cite{Gordon1973}. Thus,
in the low-frequency limit the Minkowski momentum should be identified with
Gordon's pseudo-momentum,\ or in the language of this paper, with the canonical\ momentum.

Most of the previous treatments of the Minkowski momentum have been restricted
to nondispersive media \cite{Brevik2004}, so an alternative procedure would be
to interpret the canonical momentum as the appropriate generalization of the
Minkowski momentum to dispersive media. In this approach, the definition
(\ref{MinkowskiTotMom2}) of the Minkowski momentum would be dropped and
replaced by the classical limit of the definition (\ref{momdis}) of the
canonical momentum.

Similar results follow from the alternative classical expressions of the total
angular momentum. The classical angular momentum defined by the Abraham
momentum density is%
\begin{equation}
\boldsymbol{\mathcal{J}} _{A}=\int d^{3}r\ \ \mathbf{r\times g}_{A}\left(
\mathbf{r},t\right)  ,
\end{equation}
so the corresponding quantum operator is%
\begin{equation}
\mathbf{J}_{A}=\int\frac{d^{3}k}{\left(  2\pi\right)  ^{3}}\frac{v_{gr}\left(
k\right)  v_{ph}\left(  k\right)  }{c^{2}}\left\{  \mathbf{a}_{j}^{\dagger
}\left(  \mathbf{k}\right)  \left(  \hslash\mathbf{k\times}\frac{1}{i}%
\frac{\partial}{\partial\mathbf{k}}\right)  _{i}\mathbf{a}_{j}\left(
\mathbf{k}\right)  +\frac{\mathbf{k}}{k}\sum_{s}\hslash sa_{s}^{\dagger
}\left(  \mathbf{k}\right)  a_{s}\left(  \mathbf{k}\right)  \right\}  .
\end{equation}
Similarly the Minkowski angular momentum%
\begin{equation}
\boldsymbol{\mathcal{J}} _{M}=\int d^{3}r\ \ \mathbf{r\times g}_{M}\left(
\mathbf{r},t\right)
\end{equation}
leads to the operator%
\begin{equation}
\mathbf{J}_{M}=\int\frac{d^{3}k}{\left(  2\pi\right)  ^{3}}\frac{v_{gr}\left(
k\right)  }{v_{ph}\left(  k\right)  }\left\{  \mathbf{a}_{j}^{\dagger}\left(
\mathbf{k}\right)  \left(  \hslash\mathbf{k\times}\frac{1}{i}\frac{\partial
}{\partial\mathbf{k}}\right)  _{i}\mathbf{a}_{j}\left(  \mathbf{k}\right)
+\frac{\mathbf{k}}{k}\sum_{s}\hslash sa_{s}^{\dagger}\left(  \mathbf{k}%
\right)  a_{s}\left(  \mathbf{k}\right)  \right\}  .
\end{equation}

\section{\label{&experiment}Experimental tests}

\subsection{Radiation-pressure experiment of Jones and Leslie}

An important experiment which bears on the question of the momentum of light
in dielectric media was carried out by Jones and Leslie
\cite{Jones-Leslie1978}. In this work the radiation pressure of a light beam
striking a mirror immersed in various optically dense liquids was measured
with high accuracy. Each measurement was compared to the radiation pressure of
the same light beam striking the same mirror in air. The experimental data
showed that the mechanical momentum imparted to the mirror is directly
proportional to the index of refraction $n\left(  \omega\right)  $ of the
medium to within $\pm0.05\%$. Several alternative hypotheses, such as
proportionality to the \textquotedblleft group index \textquotedblright\
\begin{equation}
n_{gr}\left(  \omega\right)  =n\left(  \omega\right)  +\omega dn/d\omega
\label{group-index}%
\end{equation}
or inverse proportionality to $n\left(  \omega\right)  $, were excluded by
many standard deviations.

At the heart of this experiment is a \textquotedblleft radiation-pressure
mirror,\textquotedblright\ fabricated from multilayer dielectric coatings with
high reflectivity and low absorption at the $632.8\
\operatorname{nm}%
$ wavelength of the helium-neon laser used in the experiment. This mirror is
located near the bottom of the apparatus, where it is attached by epoxy to a
thin, central vertical wire. The mirror and wire can be immersed in a variety
of dielectric liquids. A high-intensity, $15$\ $%
\operatorname{mW}%
$ helium-neon laser beam is directed near normal incidence towards this lower
mirror, and the radiation pressure exerted by the laser beam generates a
torque upon the wire. In the experiment, the resulting torque is measured both
before and after a dielectric liquid is poured into the space surrounding the mirror.

A second, \textquotedblleft twist-detecting\textquotedblright\ mirror (called
an \textquotedblleft optical lever\textquotedblright) is attached to the same
wire near the top of the apparatus, and is also immersed in the liquid. In
this way, the central wire connecting the two mirrors transmits the mechanical
torque generated by the radiation pressure from the lower to the upper mirror.
The wire is wrapped around the upper mirror many times so as to form a
current-carrying coil which, in the presence of a uniform magnetic field,
exerts a torque on the upper mirror. The reflected light signal from the upper
mirror is detected by a pair of balanced photodiodes, and is used as the
primary input into a feedback circuit that controls the current in the coil,
so that the torque generated by its interaction with the magnetic field
exactly cancels the torque arising from the radiation pressure exerted by the
laser beam on the lower mirror. (The radiation pressure exerted upon the upper
mirror by the low-intensity light beam for monitoring the angular displacement
of the \textquotedblleft optical lever\textquotedblright\ is negligible.) The
central wire is grounded at the bottom of the metallic apparatus, and is
insulated from the top, in order for a current to be fed through the wire.

The use of a counterbalancing torque generated in the upper mirror guarantees
that no mechanical motion of the lower mirror, or of the fluid, ever occurs
during a measurement, i.e., these are \emph{null measurements}. Nonlinearities
in the system do not affect the position of the null, and also there is no
need to include any hydrodynamic effects (including electrostrictive pressure
effects) in the calculation of the radiation pressure. After the system has
been balanced and comes into mechanical equilibrium, a measurement of the
current passing through the coil around the upper mirror is a direct measure
of the radiation pressure exerted by the laser beam on the lower mirror.

The experiment employs synchronous detection to cancel out systematic errors.
The laser beam is periodically translated from the left side to the right side
of the radiation-pressure mirror with respect to the central wire. This is
done symmetrically, so that the radiation-pressure-generated torque
periodically reverses sign. The electronic feedback system is designed so that
the current sent to the coil wrapped around the upper mirror is also reversed
in sign in synchronism with the periodic switching of the laser beam.
Derivative feedback to the coil around the upper mirror is used to achieve
critical damping of this torsional-oscillator system.

We will analyze this experiment by assuming that each photon in the beam
carries momentum $\mathbf{p}$ that is normal to the mirror. Let us call the
rate of arrival of photons normally incident at the mirror $\dot{N}_{inc}$.
For a perfectly reflective mirror the momentum transfer per photon at normal
incidence is $2\mathbf{p}$, so the magnitude $F_{rad}=\left\vert
\mathbf{F}_{rad}\right\vert $ of the force due to the flux of photons striking
the mirror at normal incidence is%
\begin{equation}
F_{rad}=\dot{N}_{inc}2\left\vert \mathbf{p}\right\vert
\end{equation}
The entrance window to the apparatus is anti-reflection coated, and there is
negligible absorption in the liquid; therefore the rate of arrival of laser
photons at the mirror is the same as the rate of arrival of laser photons at
the entrance window. If the entire laser output is focussed through the
entrance window onto the surface of the mirror, $\dot{N}_{inc}$ is closely
approximated by%
\begin{equation}
\dot{N}_{inc}=\frac{P_{laser}}{\hbar\omega_{L}}\text{ ,}%
\end{equation}
where $P_{laser}$ is the output power of the laser and $\hbar\omega_{L}$ is
the energy per laser photon.

There are three possible choices for $\mathbf{p}$. For $\mathbf{p}%
=\mathbf{p}_{can}=\hslash\mathbf{k}$, the force on the mirror is%
\begin{equation}
\left(  F_{rad}\right)  _{can}=\dot{N}_{inc}2\hslash k=n\left(  \omega
_{L}\right)  \left[  \dot{N}_{inc}2\frac{\hslash\omega_{L}}{c}\right]
=n\left(  \omega_{L}\right)  \left[  2\frac{P_{laser}}{c}\right]  ,
\end{equation}
where we have used the dispersion relation $k=n\left(  \omega_{L}\right)
\omega_{L}/c$. For $\mathbf{p}=\mathbf{p}_{M}$ or $\mathbf{p}=\mathbf{p}_{A} $
the relations (\ref{MinkowskiPhotonMom}) and (\ref{AbrahamPhotonMom}) yield
the corresponding forces%
\begin{equation}
\left(  F_{rad}\right)  _{M}=\frac{n^{2}}{n_{gr}}\left[  2\frac{P_{laser}}%
{c}\right]  ,
\end{equation}
and%
\begin{equation}
\left(  F_{rad}\right)  _{A}=\frac{1}{n_{gr}}\left[  2\frac{P_{laser}}%
{c}\right]  ,
\end{equation}
where $n_{gr}$ is the group index defined in (\ref{group-index}).

In each case we want to calculate the ratio%
\begin{equation}
R=\frac{F_{rad}\left(  \text{dielectric}\right)  }{F_{rad}\left(
\text{air}\right)  }%
\end{equation}
of the radiation-pressure forces on the mirror with and without the liquid.
Since $n=n_{gr}=1$ in air, the three alternative values are
\begin{equation}
R_{can}=n\left(  \omega_{L}\right)  , \label{CanonicalRatio}%
\end{equation}%
\begin{equation}
R_{M}=\frac{n^{2}\left(  \omega_{L}\right)  }{n_{gr}\left(  \omega_{L}\right)
}, \label{MinkowskiRatio}%
\end{equation}
and%
\begin{equation}
R_{A}=\frac{1}{n_{gr}\left(  \omega_{L}\right)  }, \label{AbrahamRatio}%
\end{equation}
for the canonical, Minkowski, and Abraham momenta, respectively.

The results of evaluating the alternative values (\ref{CanonicalRatio}%
)-(\ref{AbrahamRatio}) of the ratio $R$ using the data provided by Jones and
Leslie are presented in Table \ref{!ourtable}. For each dielectric we show the
average experimental value $R_{\exp}$ and the corresponding standard deviation
$\sigma$, together with the predicted values and their differences from the
experimental value expressed as a multiple of $\sigma$. For example, in the
case of benzene the observed ratio differs from the Minkowski prediction
(\ref{MinkowskiRatio}) by $22$ standard deviations and from the Abraham
prediction (\ref{AbrahamRatio}) by $405$ standard deviations!%

\begin{table}[hbt] \centering
\begin{tabular}
[c]{|l|l|l|l|l|}\hline
\textbf{liquid} & $R_{\exp}$ & $R_{can}$ & $R_{M}$ & $R_{A}$\\\hline
methanol & $%
\begin{array}
[c]{c}%
\mathbf{1.3281\pm\sigma}\\
\left(  \sigma=0.0018\right)
\end{array}
$ & $%
\begin{array}
[c]{c}%
1.3275\\
\left(  -0.3\sigma\right)
\end{array}
$ & $%
\begin{array}
[c]{c}%
1.3134\\
\left(  -8.2\sigma\right)
\end{array}
$ & $%
\begin{array}
[c]{c}%
0.7453\\
\left(  -324\sigma\right)
\end{array}
$\\\hline
acetone & $%
\begin{array}
[c]{c}%
\mathbf{1.3553\pm\sigma}\\
\left(  \sigma=0.0018\right)
\end{array}
$ & $%
\begin{array}
[c]{c}%
1.3563\\
\left(  +0.6\sigma\right)
\end{array}
$ & $%
\begin{array}
[c]{c}%
1.3359\\
\left(  -10.8\sigma\right)
\end{array}
$ & $%
\begin{array}
[c]{c}%
0.7262\\
\left(  -350\sigma\right)
\end{array}
$\\\hline
ethanol & $%
\begin{array}
[c]{c}%
\mathbf{1.3594\pm\sigma}\\
\left(  \sigma=0.0022\right)
\end{array}
$ & $%
\begin{array}
[c]{c}%
1.3606\\
\left(  +0.5\sigma\right)
\end{array}
$ & $%
\begin{array}
[c]{c}%
1.3437\\
\left(  -7.1\sigma\right)
\end{array}
$ & $%
\begin{array}
[c]{c}%
0.7259\\
\left(  -288\sigma\right)
\end{array}
$\\\hline
isopropanol & $%
\begin{array}
[c]{c}%
\mathbf{1.3762\pm\sigma}\\
\left(  \sigma=0.0020\right)
\end{array}
$ & $%
\begin{array}
[c]{c}%
1.3756\\
\left(  -0.3\sigma\right)
\end{array}
$ & $%
\begin{array}
[c]{c}%
1.3577\\
\left(  -9.3\sigma\right)
\end{array}
$ & $%
\begin{array}
[c]{c}%
0.7175\\
\left(  -329\sigma\right)
\end{array}
$\\\hline
CCl$_{4}$ & $%
\begin{array}
[c]{c}%
\mathbf{1.4614\pm\sigma}\\
\left(  \sigma=0.0021\right)
\end{array}
$ & $%
\begin{array}
[c]{c}%
1.4581\\
\left(  -1.6\sigma\right)
\end{array}
$ & $%
\begin{array}
[c]{c}%
1.4313\\
\left(  -14.3\sigma\right)
\end{array}
$ & $%
\begin{array}
[c]{c}%
0.6732\\
\left(  -375\sigma\right)
\end{array}
$\\\hline
toluene & $%
\begin{array}
[c]{c}%
\mathbf{1.4898\pm\sigma}\\
\left(  \sigma=0.0018\right)
\end{array}
$ & $%
\begin{array}
[c]{c}%
1.4921\\
\left(  +1.3\sigma\right)
\end{array}
$ & $%
\begin{array}
[c]{c}%
1.4528\\
\left(  -20.5\sigma\right)
\end{array}
$ & $%
\begin{array}
[c]{c}%
0.6525\\
\left(  -465\sigma\right)
\end{array}
$\\\hline
benzene & $%
\begin{array}
[c]{c}%
\mathbf{1.4970\pm\sigma}\\
\left(  \sigma=0.0021\right)
\end{array}
$ & $%
\begin{array}
[c]{c}%
1.4974\\
\left(  +0.2\sigma\right)
\end{array}
$ & $%
\begin{array}
[c]{c}%
1.4518\\
\left(  -21.5\sigma\right)
\end{array}
$ & $%
\begin{array}
[c]{c}%
0.6475\\
\left(  -405\sigma\right)
\end{array}
$\\\hline
\end{tabular}
\caption{Ratios of radiation pressure in liquid to that in air (data from
\cite{Jones-Leslie1978}).} \label{!ourtable}%
\end{table}%

Therefore the Jones and Leslie experiment demonstrates that near normal
incidence the radiation pressure on a mirror immersed in a dielectric liquid
is given by the rate of transfer of the \emph{canonical} momentum
$\hslash\mathbf{k}$ per photon within an accuracy of $\pm0.05\%$. In this
connection it is important to note that the theories of Gordon
\cite{Gordon1973} and Loudon \cite{Loudon2002} both predict that the radiation
pressure force on a mirror immersed in a dispersionless dielectric will be
determined by the Minkowski, rather than the Abraham, momentum. As we have
noted above, the Minkowski and canonical momenta agree for dispersionless
materials, but we have further demonstrated in Table \ref{!ourtable} that the
experimental results for optical frequency radiation in a dispersive medium
decisively favor the canonical momentum over the Minkowski momentum, as well
as over the Abraham momentum.

\subsection{\label{&AbrahamExpt}Experimental relevance of the Abraham
momentum}

Of the three momenta we have studied, only the canonical momentum is required
to explain atomic recoil in spontaneous emission, the Cerenkov and Doppler
effects, and all conventional nonlinear and quantum optics experiments
involving the phase-matching relations. In addition, the radiation-pressure
experiment of Jones and Leslie is consistent with the choice of the canonical
momentum for the dressed photons. When, if ever, are the Abraham or Minkowski
forms of momentum needed? In this connection, there have been important
experiments demonstrating the relevance of the Abraham momentum by James
\cite{James1968} and by Walker \emph{et al}. \cite{Walker1977}. (For a review
of these experiments, see Brevik \cite{Brevik1979}.) These experiments, which
were first proposed by Marx and Gy\"{o}rgyi \cite{Gyorgi1955}, involve
toroidal or annular, dielectric-filled regions subjected to crossed electric
and magnetic fields, with low-frequency time variations. In particular, in the
experiment of Walker \emph{et al}., the Abraham force due to the time-varying
polarization current crossed into the magnetic field was verified to within an
accuracy of $\pm5\%$. This implies that the Minkowski theory is in
disagreement with the experimental data of Walker \emph{et al}., by 20
standard deviations.

Note that these toroidal experiments involved \textquotedblleft
closed\textquotedblright\ systems, in the sense that the dielectric medium and
electromagnetic fields are entirely enclosed, for example, within the toroidal
torsional bob of the torsional oscillator used by Walker \emph{et al}. Thus in
these experiments the dielectric medium experiences \emph{accelerated} motion
during measurements. No external forces are present, and the whole enclosed
system of fields and dielectric rotates together as a rigid body. By contrast,
the Jones and Leslie configuration involves an \textquotedblleft
open\textquotedblright\ system, in which\emph{\ }an external torque is used in
feedback to prevent any accelerated motions of the mirror and the dielectric
liquid during measurements.

Furthermore, two papers by Lai \cite{Lai1979,Lai1981} have convincingly
demonstrated theoretically that in the low-frequency or static limit, the
Minkowski momentum density would give unphysical results for the measurement
of the total angular momentum in all such closed-system experiments in which
acceleration of the dielectric is allowed. Thus the experiments by James and
by Walker \emph{et al}., and the papers by Lai, all provide strong evidence
that the Abraham, rather than the Minkowski momentum, is required for a
correct description of all such closed systems that undergo accelerated
motions. This is consistent with Planck's law of inertia for electromagnetic
energy. Since the canonical momentum is identical to the Minkowski momentum in
the static limit, these results also rule out the canonical momentum as being
physically relevant in these kinds of experiments. However, one of the
assumptions of the Milonni theory is that center-of-mass motions of atoms of
the medium are not allowed. Hence it is not surprising the canonical momentum
derived from this theory does not apply to these experiments.

\section{Conclusions}

The \emph{ad hoc} quantization scheme employed above leads in a natural way to
several forms of momentum for the electromagnetic field in a dispersive
medium. The first is the canonical momentum which is uniquely defined as the
generator of spatial translations. The conservation law for the canonical
momentum is validated by the atomic recoil in spontaneous emission, the
Cerenkov and Doppler effects, and the phase-matching conditions in nonlinear
optics. Furthermore, the canonical momentum correctly predicts the results of
the Jones and Leslie radiation-pressure experiment. The explicit appearance of
the group velocity in the \emph{ad hoc} scheme suggests that experiments to
measure quantum fluctuations of the electromagnetic field in a variety of
dielectric media would be of great interest.

The second form, the kinetic momentum, is not unique, since the operators are
derived by quantizing the classical expressions of the Abraham and Minkowski
momenta. The experiments discussed in Section V-B demonstrate the experimental
relevance of the Abraham, as opposed to the canonical, momentum for closed
systems. Since these experiments have all been carried out for classical,
low-frequency fields, they do not provide direct evidence for the meaning of
the operators $\mathbf{P}_{A}$ or $\mathbf{P}_{M}$. Investigating the quantum
significance of the Abraham or Minkowski momenta would again require
experiments sensitive to quantum fluctuations.

In addition to these experimental questions, there are also issues of
theoretical consistency that have to be faced. The conjectured form
(\ref{m-eff-A}) of the Abraham effective photon mass is based on the implicit
assumption that the dressed photon model can be applied to accelerated media.
This is inconsistent with the basic assumption in the quantization scheme that
no center-of-mass acceleration of the atoms occurs. One possible way to
resolve this contradiction would be to imitate Milonni's scheme by starting
with a classical expression for the electromagnetic energy in an accelerated medium.

\begin{acknowledgement}
\begin{acknowledgments}
\begin{acknowledgement}
It is a pleasure to acknowledge stimulating discussions with Iver Brevik, J.
David Jackson, Ulf Leonhardt, Peter Milonni, Rodney Loudon, and Achilles
Speliotopoulos. This work was supported in part by the NSF.
\end{acknowledgement}
\end{acknowledgments}
\end{acknowledgement}

\end{document}